\documentclass[prl,aps,twocolumn,tightenlines,notitlepage,nofootinbib,preprintnumbers,letterpaper,superscriptaddress]{revtex4-2} 

\pdfoutput=1
\usepackage{graphicx,mathtools,tabu}
\usepackage{epstopdf}
\usepackage{amsmath, bm}
\usepackage{amsfonts}
\usepackage[colorlinks=true,breaklinks=true,citecolor=magenta]{hyperref}
\usepackage{accents}
\usepackage[figure, table]{hypcap}
\usepackage{amssymb}
\usepackage{appendix}
\usepackage{comment}
\usepackage{bbold}
\usepackage{xcolor}
\usepackage{color}
\usepackage{slashed}
\usepackage{subfigure}
\usepackage{setspace}
\usepackage{footnote}
\usepackage{multirow}
\usepackage{braket}
\usepackage[normalem]{ulem}
\usepackage[utf8]{inputenc}
\usepackage{mathrsfs}
\usepackage{longtable}
\usepackage{enumitem}
\usepackage{cleveref}

\usepackage{siunitx}
\DeclareSIUnit \s {\second}
\DeclareSIUnit \ns {\nano\second}
\DeclareSIUnit \mus {\micro\second}
\DeclareSIUnit \ms {\milli\second}
\DeclareSIUnit \MB {\mega\byte}
\DeclareSIUnit \GB {\giga\byte}
\DeclareSIUnit \TB {\tera\byte}
\DeclareSIUnit \PB {\peta\byte}
\DeclareSIUnit \Mbps {\mega\bit/\s}
\DeclareSIUnit \Gbps {\giga\bit/\s}
\DeclareSIUnit \Tbps {\tera\bit/\s}
\DeclareSIUnit \Pbps {\peta\bit/\s}
\DeclareSIUnit \kton {\kilo\tonne} % changed  back to kton
\DeclareSIUnit \kt {\kilo\tonne}
\DeclareSIUnit \kty {\kilo\tonne-\year}
\DeclareSIUnit \Mt {\mega\tonne}
\DeclareSIUnit \eV {\electronvolt}
\DeclareSIUnit \keV {\kilo\electronvolt}
\DeclareSIUnit \MeV {\mega\electronvolt}
\DeclareSIUnit \GeV {\giga\electronvolt}
\DeclareSIUnit \TeV {\tera\electronvolt}
\DeclareSIUnit \PeV {\peta\electronvolt}
\DeclareSIUnit \EeV {\exa\electronvolt}
\DeclareSIUnit \sr {sr}
\DeclareSIUnit \m {\meter}
\DeclareSIUnit \cm {\centi\meter}
\DeclareSIUnit \nm {\nano\meter}
\DeclareSIUnit \in {\inchcommand}
\DeclareSIUnit \km {\kilo\meter}
\DeclareSIUnit \kV {\kilo\volt}
\DeclareSIUnit \kW {\kilo\watt}
\DeclareSIUnit \MW {\mega\watt}
\DeclareSIUnit \MHz {\mega\hertz}
\DeclareSIUnit \mrad {\milli\radian}
\DeclareSIUnit \year {years}
\DeclareSIUnit \POT {POT}
\DeclareSIUnit \sig {$\sigma$}
\DeclareSIUnit\parsec{pc}
\DeclareSIUnit\lightyear{ly}
\DeclareSIUnit\foot{ft}
\DeclareSIUnit\ft{ft}
\DeclareSIUnit \ppb{ppb}
\DeclareSIUnit \ppt{ppt}
\DeclareSIUnit \samples{S}
\DeclareSIUnit \pe{PE}
\DeclareSIUnit \GeVmwe{GeV/mwe}
\DeclareSIUnit \mwe{mwe}

\newcommand{\enu}{\E_\enu}

\begin{document}

\def\aj{\ref@jnl{AJ}}                   % Astronomical Journal
\def\actaa{\ref@jnl{Acta Astron.}}      % Acta Astronomica
\def\araa{\ref@jnl{ARA\&A}}             % Annual Review of Astron and Astrophys
\def\apj{\ref@jnl{ApJ}}                 % Astrophysical Journal
\def\apjl{\ref@jnl{ApJ}}                % Astrophysical Journal, Letters
\def\apjs{\ref@jnl{ApJS}}               % Astrophysical Journal, Supplement
\def\ao{\ref@jnl{Appl.~Opt.}}           % Applied Optics
\def\apss{\ref@jnl{Ap\&SS}}             % Astrophysics and Space Science
\def\aap{\ref@jnl{A\&A}}                % Astronomy and Astrophysics
\def\aapr{\ref@jnl{A\&A~Rev.}}          % Astronomy and Astrophysics Reviews
\def\aaps{\ref@jnl{A\&AS}}              % Astronomy and Astrophysics, Supplement
\def\azh{\ref@jnl{AZh}}                 % Astronomicheskii Zhurnal
\def\baas{\ref@jnl{BAAS}}               % Bulletin of the AAS
\def\bac{\ref@jnl{Bull. astr. Inst. Czechosl.}}
                % Bulletin of the Astronomical Institutes of Czechoslovakia 
\def\caa{\ref@jnl{Chinese Astron. Astrophys.}}
                % Chinese Astronomy and Astrophysics
\def\cjaa{\ref@jnl{Chinese J. Astron. Astrophys.}}
                % Chinese Journal of Astronomy and Astrophysics
\def\icarus{\ref@jnl{Icarus}}           % Icarus
\def\jcap{\ref@jnl{J. Cosmology Astropart. Phys.}}
                % Journal of Cosmology and Astroparticle Physics
\def\jrasc{\ref@jnl{JRASC}}             % Journal of the RAS of Canada
\def\memras{\ref@jnl{MmRAS}}            % Memoirs of the RAS
\def\mnras{\ref@jnl{MNRAS}}             % Monthly Notices of the RAS
\def\na{\ref@jnl{New A}}                % New Astronomy
\def\nar{\ref@jnl{New A Rev.}}          % New Astronomy Review
\def\pra{\ref@jnl{Phys.~Rev.~A}}        % Physical Review A: General Physics
\def\prb{\ref@jnl{Phys.~Rev.~B}}        % Physical Review B: Solid State
\def\prc{\ref@jnl{Phys.~Rev.~C}}        % Physical Review C
\def\prd{\ref@jnl{Phys.~Rev.~D}}        % Physical Review D
\def\pre{\ref@jnl{Phys.~Rev.~E}}        % Physical Review E
\def\prl{\ref@jnl{Phys.~Rev.~Lett.}}    % Physical Review Letters
\def\pasa{\ref@jnl{PASA}}               % Publications of the Astron. Soc. of Australia
\def\pasp{\ref@jnl{PASP}}               % Publications of the ASP
\def\pasj{\ref@jnl{PASJ}}               % Publications of the ASJ
\def\rmxaa{\ref@jnl{Rev. Mexicana Astron. Astrofis.}}%
                % Revista Mexicana de Astronomia y Astrofisica
\def\qjras{\ref@jnl{QJRAS}}             % Quarterly Journal of the RAS
\def\skytel{\ref@jnl{S\&T}}             % Sky and Telescope
\def\solphys{\ref@jnl{Sol.~Phys.}}      % Solar Physics
\def\sovast{\ref@jnl{Soviet~Ast.}}      % Soviet Astronomy
\def\ssr{\ref@jnl{Space~Sci.~Rev.}}     % Space Science Reviews
\def\zap{\ref@jnl{ZAp}}                 % Zeitschrift fuer Astrophysik
\def\nat{\ref@jnl{Nature}}              % Nature
\def\iaucirc{\ref@jnl{IAU~Circ.}}       % IAU Cirulars
\def\aplett{\ref@jnl{Astrophys.~Lett.}} % Astrophysics Letters
\def\apspr{\ref@jnl{Astrophys.~Space~Phys.~Res.}}
                % Astrophysics Space Physics Research
\def\bain{\ref@jnl{Bull.~Astron.~Inst.~Netherlands}} 
                % Bulletin Astronomical Institute of the Netherlands
\def\fcp{\ref@jnl{Fund.~Cosmic~Phys.}}  % Fundamental Cosmic Physics
\def\gca{\ref@jnl{Geochim.~Cosmochim.~Acta}}   % Geochimica Cosmochimica Acta
\def\grl{\ref@jnl{Geophys.~Res.~Lett.}} % Geophysics Research Letters
\def\jcp{\ref@jnl{J.~Chem.~Phys.}}      % Journal of Chemical Physics
\def\jgr{\ref@jnl{J.~Geophys.~Res.}}    % Journal of Geophysics Research
\def\jqsrt{\ref@jnl{J.~Quant.~Spec.~Radiat.~Transf.}}
                % Journal of Quantitiative Spectroscopy and Radiative Transfer
\def\memsai{\ref@jnl{Mem.~Soc.~Astron.~Italiana}}
                % Mem. Societa Astronomica Italiana
\def\nphysa{\ref@jnl{Nucl.~Phys.~A}}   % Nuclear Physics A
\def\physrep{\ref@jnl{Phys.~Rep.}}   % Physics Reports
\def\physscr{\ref@jnl{Phys.~Scr}}   % Physica Scripta
\def\planss{\ref@jnl{Planet.~Space~Sci.}}   % Planetary Space Science
\def\procspie{\ref@jnl{Proc.~SPIE}}   % Proceedings of the SPIE

\let\astap=\aap
\let\apjlett=\apjl
\let\apjsupp=\apjs
\let\applopt=\ao

\title{Probing Pseudo-Dirac Neutrinos with Astrophysical Sources at IceCube}

\author{Kiara~Carloni}
\email{kcarloni@g.harvard.edu}
\affiliation{Department of Physics \& Laboratory for Particle Physics and Cosmology, Harvard University, Cambridge, MA 02138, USA}

\author{Ivan~Mart\'inez-Soler}
\email{imartinezsoler@fas.harvard.edu}
\affiliation{Department of Physics \& Laboratory for Particle Physics and Cosmology, Harvard University, Cambridge, MA 02138, USA}

\author{Carlos~A.~Arg{\"u}elles}
\email{carguelles@fas.harvard.edu}
\affiliation{Department of Physics \& Laboratory for Particle Physics and Cosmology, Harvard University, Cambridge, MA 02138, USA}

\author{K. S.~Babu}
\email{kaladi.babu@okstate.edu}
\affiliation{Department of Physics, Oklahoma State University, Stillwater, OK 74078, USA}

\author{P.~S.~Bhupal~Dev}
\email{bdev@wustl.edu}
\affiliation{Department of Physics and McDonnell Center for the Space Sciences, Washington University, St.~Louis, MO 63130, USA}

\date{\today}

\begin{abstract}
The recent observation of NGC~1068 by the IceCube Neutrino Observatory has opened a new window to neutrino physics with  astrophysical baselines.
In this \textit{Letter}, we propose a new method to probe the nature of neutrino masses using these observations. In particular, our method enables searching for signatures of pseudo-Dirac neutrinos with mass-squared differences that reach down to $\delta m^2 \gtrsim 10^{-21}\si\eV^2$, improving the reach of terrestrial experiments by more than a billion.
Finally, we discuss how the discovery of a constellation of neutrino sources can further increase the sensitivity and cover a wider range of $\delta m^2$ values.
\end{abstract}

\maketitle

\textbf{\emph{Introduction.---}}
Since the beginning of time, humans have stared at the sky and wondered about the universe.
Through careful inspection, we discovered the patterns that rule the motions of planets, and followed a trail of questioning that led to the theory of general relativity.
Now, equipped with enormous telescopes and modern particle physics, we can, for the first time, study tinier, even more elusive astrophysical signals.
In this \textit{Letter}, we show that these observations can be used to uncover the origin and nature of the neutrino mass.

Recently, IceCube announced the observation of the first steady-state astrophysical neutrino source, the active galactic nucleus NGC~1068~\cite{IceCube:2022der,Wright:2022vmi}. 
Assuming only that the neutrinos produced by this source follow a power-law distribution in energy, they performed a likelihood analysis and found that $79_{-20}^{+22}$ events originated from NGC~1068, yielding a rejection of the background-only hypothesis with a local (global) significance of 5.2 (4.2)~$\sigma$~\cite{IceCube:2022der}.
Neutrinos that travel to Earth from sources like NGC~1068 must traverse megaparsecs (Mpc), a distance many orders of magnitude greater than that traveled by any solar, atmospheric, reactor- or accelerator-based neutrino ever detected. 
Therefore, exploring the properties of the neutrino events from extra-galactic sources will allow us to study, for the first time, a whole class of new physics scenarios whose signals appear only at extremely long length scales.

\begin{figure}[htb]
  \centering
  \includegraphics[width=0.5\textwidth]{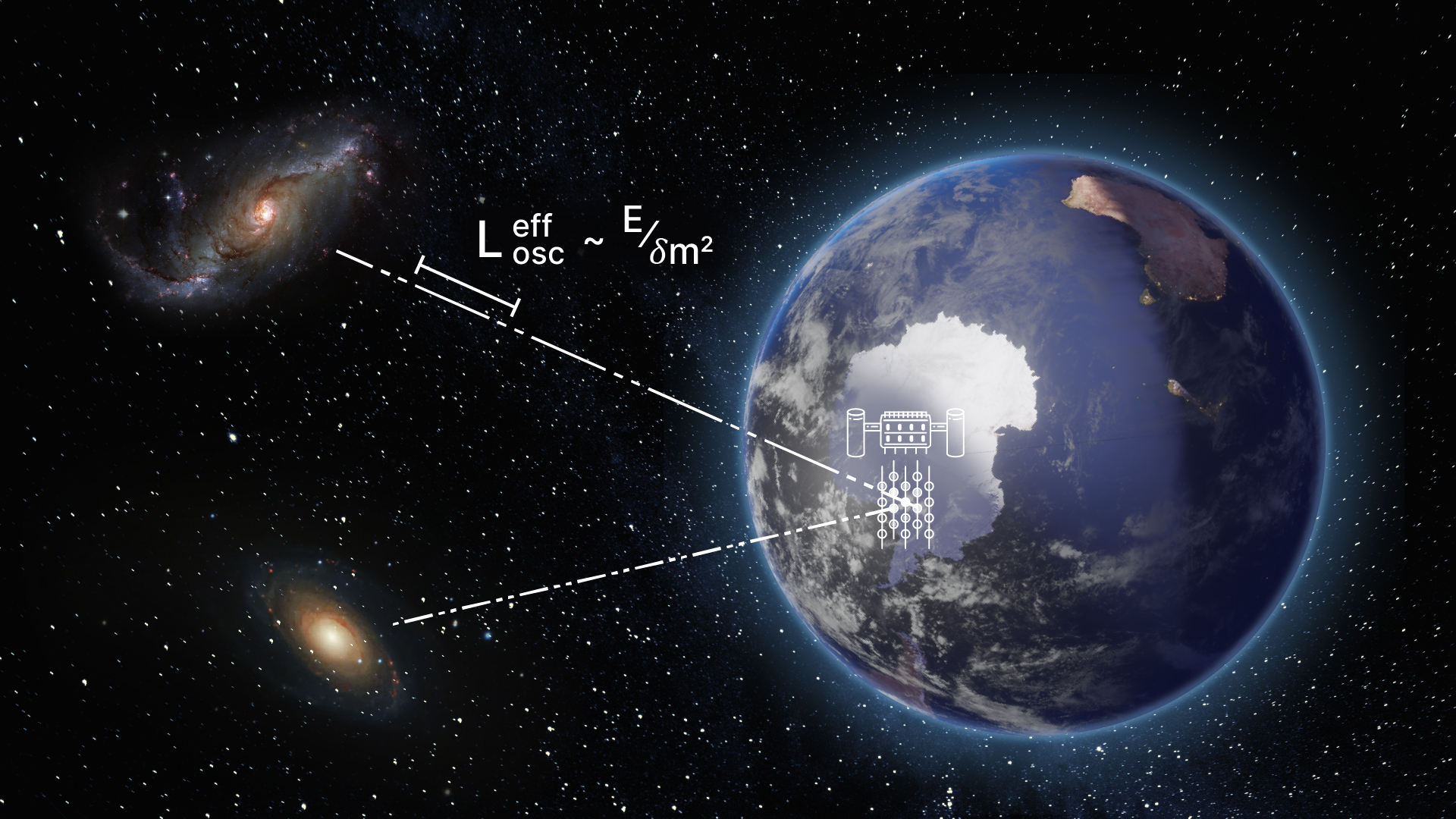}
  \caption{An artistic rendering of neutrino propagation from extra-galactic sources. Oscillation from active to sterile is depicted by the transition from solid to dashed line. }\label{fig:art}
\end{figure}

One significant example of such new physics is the pseudo-Dirac model of neutrino masses~\cite{Wolfenstein:1981kw,Petcov:1982ya,Valle:1983dk,Kobayashi:2000md}. 
In this class of models, the active neutrino mass states are accompanied by undetectable sterile states, whose masses are separated from the active ones by a tiny amount, generated by a small Majorana mass term. The active-sterile mass splittings induce an oscillation between the active and sterile neutrino states.
For very small Majorana masses, these oscillations are detectable only at extremely large values of the ratio $L/E$ (where $L$ is the baseline and $E$ denotes the neutrino energy). These large values are achievable only for astrophysical neutrino sources~\cite{Beacom:2003eu,Keranen:2003xd, Esmaili:2009fk,Esmaili:2012ac,Brdar:2018tce,DeGouvea:2020ang,Martinez-Soler:2021unz}.
See~\Cref{fig:art} for an artistic rendition of our main idea. 

In this {\it Letter}, we explore how the pseudo-Dirac neutrino scenario could be probed by observations of extra-galactic neutrinos. 
We find that currently identified astrophysical neutrino sources can provide new constraints on yet unexplored mass splittings. 
We also predict how upcoming measurements in current and future neutrino telescopes will increase the sensitivity to these new mass splittings. 

\textbf{\emph{Theory of pseudo-Dirac neutrinos.---}}
Whether or not the neutrino is its own antiparticle is a question that is yet to be settled.
In a wide class of beyond the Standard Model (BSM) theories, the neutrino is a Majorana fermion, which is its own antiparticle.
However, there does exist a class of theories where the neutrino is a four-component Dirac fermion.
Neutrino oscillation experiments are unable to distinguish its Majorana nature from a Dirac one.
In this context, it is also possible that the neutrino is a {\it pseudo-Dirac particle}~\cite{Wolfenstein:1981kw,Petcov:1982ya,Valle:1983dk,Kobayashi:2000md}, which is fundamentally a Majorana fermion, but essentially acts like a Dirac fermion in most experimental settings.
However, they can be differentiated through active-sterile oscillations when the baseline traversed by the neutrino is extremely long relative to the detection energy.

The mass matrix spanning the active species $\nu_a$ and its Dirac partner $\nu_s$ has the form (with multiple flavors)
\begin{eqnarray}\label{eq:seesaw}
M_\nu = \left( \begin{matrix}  0 & m_D \cr m_D^T & M_R \end{matrix} \right).
\end{eqnarray}
If $M_R = 0$ in Eq.~\eqref{eq:seesaw}, lepton number is preserved, and the neutrino is a Dirac particle; if $M_R \neq 0$, it is a Majorana particle; and if, in the eigenvalue sense, $|M_R| \ll |m_D|$, it is a pseudo-Dirac particle. 

Phenomenologically, a pseudo-Dirac neutrino is one logical possibility in the context of neutrino mass generation.
At first sight, the condition $|M_R| \ll |m_D|$ may not look natural since $M_R$ is a gauge-invariant mass term in the SM, which could be much larger than the electroweak-symmetry-breaking scale, as e.g., in the original {\it seesaw mechanism}~\cite{Minkowski:1977sc, Mohapatra:1979ia, Yanagida:1979as, GellMann:1980vs}. 
The smallness of the neutrino mass, compared to the charged fermion masses, would remain unexplained in this case of vanishing $M_R$.  
However, there are theories where  $M_D$ is naturally small and $M_R = 0$ at the renormalizable level. Nonzero elements of $M_R$ are induced via higher-dimensional operators suppressed by the inverse Planck scale.
This is the case in the {\it Dirac seesaw  scenario}~\cite{Silagadze:1995tr,Joshipura:2013yba,Gu:2006dc,Ma:2014qra,Valle:2016kyz,CentellesChulia:2018bkz}, which is realized naturally in the mirror universe model~\cite{Lee:1956qn,Foot:1991py,Berezhiani:1995yi}.
Such theories provide a better understanding of parity ($P$) violation, since $P$ is an unbroken (or spontaneously broken) symmetry in this context.
They provide mirror partners for every SM fermion, including lepton doublets $\Psi' = (\nu', \,\ell')$ of a mirror $SU(2)'_L$ symmetry which are the partners of the usual $SU(2)_L$ lepton doublet $\Psi = (\nu ,\, \ell)$.
In this context, $\nu'$ plays the role of sterile neutrinos, with its mass protected by the $SU(2)'_L$ gauge symmetry.
A Dirac mass term connecting $\nu$ and $\nu'$ would arise from a generalized seesaw mechanism.

Operators of the type $(\Psi \Psi')(H H')/M_N$, where $H$ and $H'$ are the Higgs doublets of $SU(2)_L$ and mirror $SU(2)'_L$, respectively, are induced once a heavy neutral lepton $N$ is integrated out.
Specifically, $N$ has interactions given by $(\Psi N H)+ (\Psi' N' H') + (M_N/2) N N'$.
Lepton number remains unbroken in this scenario, which also explains why the Dirac mass term $m_D = v v'/M_N$ (where $v$ and $v'$ are the vacuum expectation values of $H$ and $H'$ respectively) is very small.
Alternatively, a bi-doublet Higgs $\Phi(2,2)$ with the couplings $\Psi \Psi' \Phi + \mu H H' \Phi^*$ could lead to the same operator with a coefficient $(\mu /M_\Phi^2)$, once the $\Phi$ field is integrated out.

Now, quantum gravity corrections are expected to break all global symmetries, such as lepton number.
One would then expect dimension-5 Weinberg operators~\cite{Weinberg:1979sa} of the type $(\Psi \Psi H H)/M_{\rm Pl}$ and $(\Psi' \Psi' H' H')/M_{\rm Pl}$ would then be induced by gravity, with coefficients presumably of order unity.
This would result in small diagonal entries of $M_\nu$ in Eq.~\eqref{eq:seesaw}, implying a pseudo-Dirac neutrino.
%Such mass splittings would imply that oscillation occurs between the active and sterile states, leading to ``dips'' in the spectrum observed in neutrino detectors, which are sensitive only to the active components.
In the mirror neutrino scenario, one would expect the active-sterile mass splitting to be on the order of  $\delta m^2 \approx (2,\,0.3) \times 10^{-7} {\rm eV}^2$
(using $m_a \simeq (0.05,\,0.007)\, {\rm eV}$ for the 
larger two of the active neutrino masses with normal ordering). However, such mass splitting values are already excluded by solar neutrino data, which requires $\delta m^2 \lesssim 10^{-11}\si\eV^2$~\cite{deGouvea:2009fp}, with Ref.~\cite{Ansarifard:2022kvy} finding a small preference for $\delta m^2\simeq 1.2\times 10^{-11}~\text{eV}^2$.\footnote{There also exist bounds on $\delta m^2\lesssim 10^{-8}\si\eV^2$ from Big Bang nucleosynthesis considerations~\cite{Barbieri:1989ti,Enqvist:1990ek}.}
This difficulty can be evaded by gauging the $B-L$ symmetry, which is anomaly-free in presence of sterile neutrinos.
This gauge symmetry is spontaneously broken by a singlet scalar field, $S$, carrying two units of $B-L$ charge.
The Weinberg operators would then be modified to the form $(\Psi \Psi H H S)/M_{\rm Pl}^2$, leading to diagonal elements of $M_\nu$ on the order of $v^2 v_{BL}/M_{\rm Pl}^2$.
For the $B-L$ symmetry breaking scale of $v_{BL} = (10^4-10^{14})\,{\rm GeV}$~\footnote{$v_{BL}\lesssim 10^4$ GeV is disfavored by the LHC null results on heavy $Z'$-resonance searches, assuming coupling strength similar to the weak interaction strength~\cite{ATLAS:2019erb, CMS:2019tbu}. %$v_{BL}\gtrsim 10^{14}$ GeV would be in conflict with the solar neutrino constraint on the mass splitting~\cite{deGouvea:2009fp, Ansarifard:2022kvy}.
} this would lead to a mass splitting of order $(10^{-22}-10^{-12})\, {\rm eV}^2$.
As we show below, a significant portion of this well-motivated range of $v_{BL}$  would be probed by the high-energy neutrinos detected at IceCube. There are other models of naturally light Dirac neutrinos in the literature with tiny masses arising from quantum loop corrections, see e.g.~Refs.~\cite{Mohapatra:1987hh,Babu:1988yq,Farzan:2012sa,Ma:2016mwh,Saad:2019bqf,Jana:2019mez,Babu:2022ikf}. 
Including Planck-suppressed higher-dimensional operators, many of these models would predict pseudo-Dirac neutrinos. 
%In particular, in a class of left-right symmetric theories with a generalized seesaw mechanism to induce charged fermion masses~\cite{Davidson:1987mh}, neutrinos are pseudo-Dirac particles with their tiny masses arising from two-loop quantum corrections ~\cite{Babu:1988yq,Babu:2022ikf}. Here the active-sterile mass splittings would depend on both the $SU(2)_R$ and the $B-L$ breaking scales.

In all the pseudo-Dirac scenarios mentioned above, the mixing between active and sterile states, given by $\tan2\theta = 2m_{D}/M_{R}$, is nearly maximal due to the pseudo-Dirac condition.
The mass eigenstates of Eq.~\eqref{eq:seesaw} are $\nu_{S} = \sin\theta~\nu_{a} + \cos\theta~\nu_{s}$ and $\nu_{A} = (-i)(\cos\theta~\nu_{a} - \sin\theta~\nu_{s})$. For very large mixing angles, those states coincide with the symmetric $(\nu_{S} = (\nu_{a} + \nu_{s})/\sqrt{2})$ and anti-symmetric $(\nu_{A} = -i(\nu_{a} - \nu_{s})/\sqrt{2})$ combinations of the active and sterile neutrinos, with their mass difference being proportional to $M_{R}$.

The sterile component may have additional interactions with the SM fermions, such as of the form $\overline{\nu_s^c}\eta^+\ell_R$, where $\eta^+$ is an $SU(2)_L$-singlet charged scalar, which can be as light as $\sim \SI{100}\GeV$ and can induce a new Glashow-like resonance at IceCube~\cite{Babu:2019vff, Babu:2022fje}.
In this {\it Letter}, we will not consider this possibility and will focus on the minimal case where all the non-standard effects arise solely from the active-sterile neutrino mixing.

\textbf{\emph{Neutrino evolution on astrophysical scales.---}}
The neutrino flavor evolution is obtained by solving the Schr\"odinger equation along the neutrino trajectory.
The time dependence of each flavor state is given by 

\begin{equation}
\ket{\nu_{\alpha}(t)} = \exp{\left(-i\int^{t}_{0} \mathcal{H}_{\alpha\beta}(t') dt'\right)}\ket{\nu_{\beta}}_{0},
\end{equation}

where $\ket{\nu_{\beta}}_{0}$ corresponds to the initial flavor state.
In vacuum, the Hamiltonian describing the neutrino evolution is $\mathcal{H}(t) = U\mathcal{M}^2 U^{\ast}/E(t)$. Here $U$ stands for the lepton mixing matrix, which relates the mass and flavor eigenstates, $\ket{\nu_{\alpha}} = U^{\ast}_{\alpha i} \ket{\nu_{i}}$, and $\mathcal{M}^2 = \text{diag}(m^2_{1}, \cdots)$ is the diagonal mass-squared matrix. 
%The coherent forward elastic scattering of astrophysical neutrinos with the C$\nu$B generate a NC matter potential just for active neutrinos that is relevant at $\delta m^2\sim 10^{-20}~\text{eV}^2$ and $E\sim 10$~PeV for densities $\mathcal{O}(300)~\text{cm}^{-3}$.
%In vacuum, the Hamiltonian describing the neutrino evolution is $\mathcal{H}(t) = U\mathcal{M}^2 U^{\ast}/E(t)$ after subtracting the average momentum of all the massive states.
%Here $U$ stands for the lepton mixing matrix, which relates the mass and flavor eigenstates, $\ket{\nu_{\alpha}} = U^{\ast}_{\alpha i} \ket{\nu_{i}}$, and $\mathcal{M}^2 = \text{diag}(m^2_{1}, \cdots)$ is the diagonal mass-squared matrix.\footnote{In the canonical three-neutrino mixing scenario, $U$ is the PMNS matrix and $\mathcal{M}^2 = \text{diag}(m^2_{1}, m^2_{2}, m^2_{3})$.}
%The integral of the Hamiltonian is carried out over all possible neutrino trajectories.

In the case of extra-galactic sources, the expansion of the universe modifies the phase of the flavor state as neutrinos propagate, having an impact on the final flavor distribution if the oscillations are not averaged out.
In the case of a homogeneous and isotropic universe, the expansion is encoded  in the scale factor ($a(t)$) that depends on the redshift (z) as $1+z = a_{0}/a$, $a_{0}$ being the scale factor today ($a_{0}=1$ for a flat universe).
The expansion rate of the universe is given by the Hubble parameter $H = \dot{a}/a$, where $\dot{a}\equiv da/dt$.
As the universe expands, there is a redshift in the neutrino energy that will also affect the phase of the flavor states.
The relation between the initial ($E'_{\nu}$) and the redshifted ($E_{\nu}$) neutrino energies  is $E_{\nu} = E'_{\nu}/(1+z)$.
The time-integration of the Hamiltonian is given by
%Considering that mixing and masses are time-independent~\cite{deGouvea:2022dtw}, the time-integration of the Hamiltonian depends only on the neutrino energy, and is given by

\begin{equation}\label{eq:phase}
\int \mathcal{H}(t) dt 
=\frac{U\mathcal{M}^2U^{\ast}}{E_{\nu}}\int \frac{dz}{H(z)(1+z)^2}
\equiv \frac{U\mathcal{M}^2U^{\ast}}{E_{\nu}} L_{\text{eff}}.
\end{equation}

The relation between the Hubble parameter and the redshift is given by
\begin{equation}
 H(z) = H_{0}\sqrt{\Omega_{m}(1+z)^3 + \Omega_{\Lambda} + (1 - \Omega_{m}-\Omega_{\Lambda})(1+z)^2},
\end{equation}
where $\Omega_{m}$ and $\Omega_{\Lambda}$ are the fractions of matter and dark energy content, and $H_0$ is the present value of the Hubble constant. For those parameters, we used the best-fit value from Planck~\cite{Planck:2018vyg} results. On astrophysical scales, the phase the flavor states get depends on the universe's expansion via the effective distance ($L_{\text{eff}}$).

\begin{figure}[t!]
  \centering
  \includegraphics[width=0.49\textwidth]{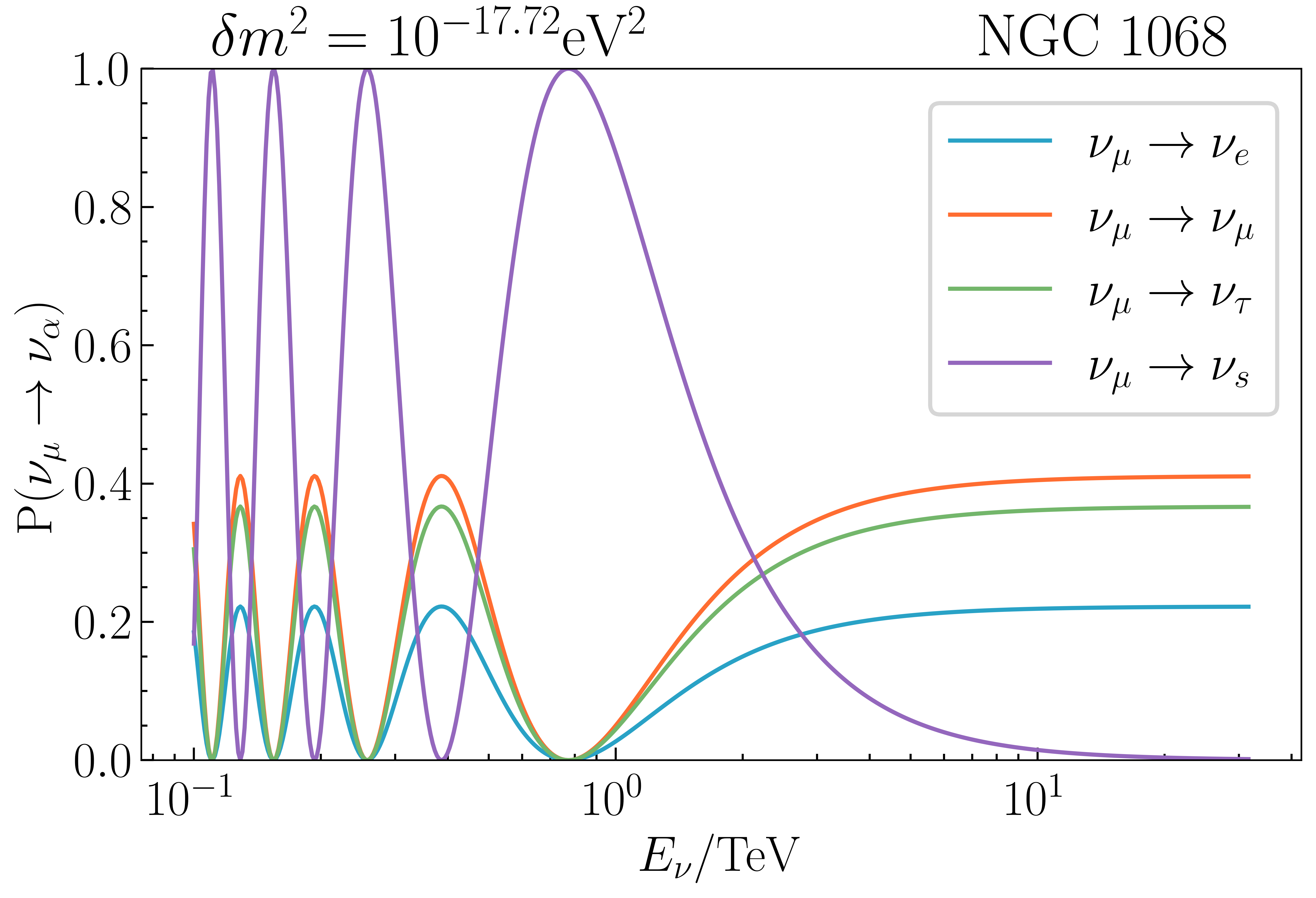}
  \caption{%\textbf{\textit{Oscillation probability}}. 
  Oscillation probability of $\nu_{\mu}$ into active and sterile components as a function of the neutrino energy for a benchmark value of $\delta m^2 = 10^{-17.72}\si\eV^2$ for all three active-sterile pairs, and for the redshift of NGC~1068 ($z=0.0038$). 
 }\label{fig:osc}
\end{figure}

In the pseudo-Dirac scenario, the mixing between the flavor and the mass eigenstates is $\nu_{\alpha} = U_{\alpha i} (\nu_{iS} + \nu_{iA})/\sqrt{2}$, where $U_{\alpha i}$ is the PMNS matrix.
Considering redshift dependence in the neutrino evolution, the probability that a flavor state $\nu_{\alpha}$ oscillates into a flavor state $\nu_{\beta}$ is given by
\begin{equation}\label{eq:phase2}
P_{\alpha\beta}= \frac{1}{4}\left|\sum_{j=1}^{3} U_{\beta j}U^{\ast}_{\alpha j}\left\{e^{\left(\frac{i m^2_{jS}L_{\text{eff}}}{2E_{\nu}}\right)}+ e^{\left(\frac{i m^2_{jA}L_{\text{eff}}}{2E_{\nu}}\right)}\right\}\right|^2
\end{equation}
where $m^2_{jA}$ and $m^2_{jS}$ are the masses of the symmetric and anti-symmetric combinations of the active and sterile states, respectively.
The oscillation probability has two well-separated oscillation lengths: For $\Delta m^2_{ij} =  m^2_{iS/A} - m^2_{jS/A} \sim 10^{-3}\si\eV^2$ (atmospheric mass splitting) or $\sim 10^{-5}\si\eV^2$ (solar mass splitting), the oscillation length is of the order of $L_{\rm osc} = 4\pi E/\Delta m^2_{ij} \sim 10^{7}-10^{9}\si\km$ for $E_{\nu}\sim \SI{10}\TeV$, which is comparable to the Earth-Sun distance.
But for the active-sterile mass splitting ($\delta m^2$), the oscillation length will be much larger, depending on the magnitude of $M_{R}$.
Taking the average over the large mass splittings, the oscillation probability becomes
\begin{equation}
P_{\alpha\beta} = \frac{1}{2}\sum^{3}_{j=1} |U_{\beta j}|^2|U_{\alpha j}|^2 \left[1 + \cos\left(\frac{\delta m^2_{j}L_{\text{eff}}}{2E_{\nu}}\right)\right].
\end{equation}
Thus, the oscillation probability depends only on the three mass splittings, one for each pair of degenerate masses.\footnote{
In the case where the mass difference is the same for the three pairs, all the flavors will oscillate with the same frequency.}
For illustration, in~\Cref{fig:osc} we show the probability of muon neutrinos oscillating into active and sterile neutrinos for all the three mass splittings equal to $\delta m^2 = 10^{-17.72}\si\eV^2$ and redshift $z=0.0038$, corresponding to NGC~1068. With regard to the lepton-mixing matrix, we used the best-fit from~\cite{Esteban:2020cvm}.  For this mass splitting, and $E_{\nu}\sim \si\TeV$, all the muon neutrinos arrive to the Earth as sterile states.

\textbf{\emph{Analysis.---}}
The discovery of high-energy extra-galactic neutrinos by the IceCube Neutrino Observatory~\cite{IceCube:2013cdw,IceCube:2013low} marked the beginning of a new era of neutrino astronomy.
According to the latest results~\cite{IceCube:2022der}, the three astrophysical sources identified with the most significance are the active galactic nuclei NGC~1068, PKS~1424+240, and TXS~0506+056, with local significance of 5.2$\sigma$, 3.7$\sigma$, and 3.5$\sigma$ respectively.
IceCube's point-source search used only track-like events, which have an excellent angular resolution~\cite{IceCube:2016tpw,IceCube:2021xar} ($\Delta \delta < 1^{\circ}$). 
In addition, they assumed that the neutrino flux followed a power-law and found that the event distribution of each source was best described by spectral indices $\hat{\gamma} = 3.2, 3.5$, and $2.0$, and total event counts $\hat{n}_s = 79, 77$, and 5, respectively; see~\Cref{table:sources}.
These sources are located at different redshifts $z=0.0038$~\cite{Meyer:2004hr}, $0.6047$~\cite{Paiano:2017pol}, and $0.3367$~\cite{Paiano:2018qeq}, corresponding to approximately 16~Mpc, 2.6~Gpc and 1.4~Gpc, respectively.  

We calculate the expected number of IceCube track-like events from each source under the standard and pseudo-Dirac hypotheses. 
To predict the expected number of events, we use the effective area given in Ref.~\cite{IceCube:2016tpw}.%\footnote{The effective area provided in Ref.~\cite{IceCube:2016tpw} yields a slightly different number of events than the one used in Ref.~\cite{IceCube:2022der} due to updates in the analysis from the earlier Ref.~\cite{IceCube:2016tpw}. To account for this discrepancy, we reschedule the number of events to match the results quoted in Ref.~\cite{IceCube:2022der}.} 
We assume that the neutrino production mechanism is charged pion decay, so the flavor composition is equal to (1:2:0) at the source.
As a benchmark scenario, we consider an initial neutrino flux following an unbroken power-law distribution in energy from $\SI{100}\GeV$ onwards, with spectral indices given by the best-fit values from Ref.~\cite{IceCube:2022der}, see~\Cref{table:sources}. 
We compute the expected number of events in a reconstructed energy bin by integrating the flux and effective area over true energy, weighted by the reconstruction probability. IceCube's energy resolution for through-going muons is about 30\% in log-energy scale, so we model this probability distribution as a Gaussian with width 0.30~\cite{IceCube:2013dkx}.  
We explore the effect of varying the energy resolution in ~\Cref{sec:energyres}. 
%We bin the resulting event expectations in energy, using bin widths motivated by IceCube's energy resolution for through-going muons, i.e. 30\% in log-scale~\cite{IceCube:2013dkx}. 
All the numerical calculations were done in \texttt{Julia}~\cite{bezanson2017julia}.

\begin{figure}[htb]
  \centering
  \includegraphics[width=0.49\textwidth]{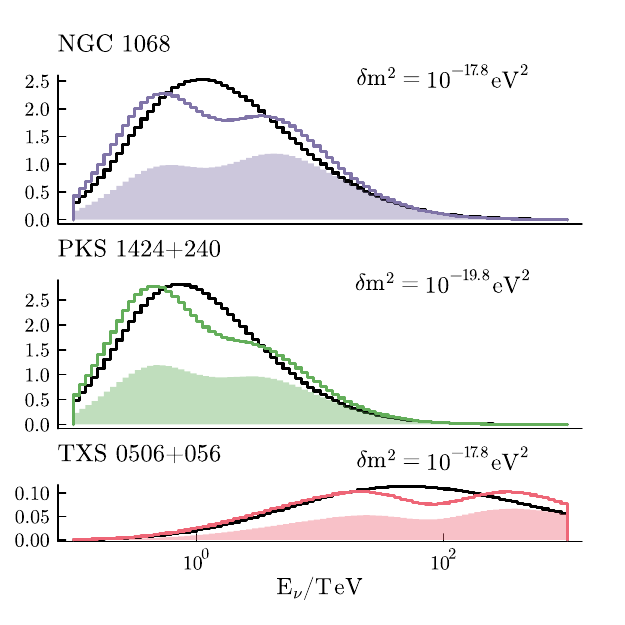}
  \caption{Calculated event distributions for the three most significant sources under the SM (black) and the pseudo-Dirac (filled color) hypotheses. The latter are plotted considering uniform sterile-active mass splittings for all three pairs of mass states; the value used in each source's calculation is printed in the upper right corner. When we maximize the likelihood that the pseudo-Dirac hypothesis can describe SM-like data, by allowing the flux parameters to vary,  (color) the difference between the two distributions is reduced.}\label{fig:event_distributions}
\end{figure}

The expected event distributions for NGC~1068, PKS~1424+240, and TXS~0506+056 for a lifetime of 3168 days are shown in~\Cref{fig:event_distributions}.
The pseudo-Dirac expectations, plotted in color, predict fewer events than the SM (black curve), since neutrinos that oscillate from active to sterile become undetectable.
%In~\Cref{fig:event_distributions} the pseudo-Dirac distributions are calculated with equal active-sterile mass-splittings for all three mass states.
Since each source has a different initial flux and a different redshift value, each is sensitive to different regions of the pseudo-Dirac parameter space.

We then calculate IceCube's sensitivity to a pseudo-Dirac signal by performing a likelihood ratio test. 
For each value of an active-sterile mass splitting, held equal over all three mass states, we calculate the Poisson likelihood of observing the SM prediction under a pseudo-Dirac hypothesis.
This procedure results in an optimistic sensitivity, since it does not account for the uncertainty in the background removal.
A slightly more conservative result could be achieved by using an effective likelihood with modeling uncertainty~\cite{Arguelles:2019izp}.
% The effective likelihood we use is a modified Poisson likelihood which incorporates a modeling uncertainty on the expectation in a given bin~\cite{Arguelles:2019izp}.
% For this analysis, the modeling uncertainty was set to 25\%, to represent the uncertainty on the total event count associated with the background-removal procedure in Ref.~\cite{IceCube:2022der}. 
We treat the flux normalization and spectral index as nuisance parameters.
% Also, we include uncertainties related to the flux normalization and the spectral index, which are considered free parameters. 
For numerical optimization, we used the package \texttt{Optim}~\cite{mogensen2018optim} in \texttt{Julia}.

Our results focus on the scenario where the three mass splittings are equal. 
This subset of the pseudo-Dirac parameter space contains the points to which we are most sensitive, because all the mass states contribute to the active-sterile oscillation at the same energies. 
Sensitivity to the scenario where two mass splittings differ from zero independently is explored in~\Cref{sec:severalmass}.

\textbf{\emph{Results.---}}
In this {\it Letter}, we perform a combined analysis of the expected event distribution of the three most significant astrophysical sources (NGC~1068, PKS~1424+240, and TXS~0506+056) observed by IceCube. 
Because the sources are unequally distant from the Earth and have different spectral indices, they are each sensitive to different regions of the $\delta m^2$ parameter space.
Combining them, we explore for the first time mass-splittings in the range $\delta m^2\in [10^{-21}, 10^{-16}]\si\eV^2$. 

\begin{figure}[htb]
  \centering
  \includegraphics[width=0.49\textwidth]{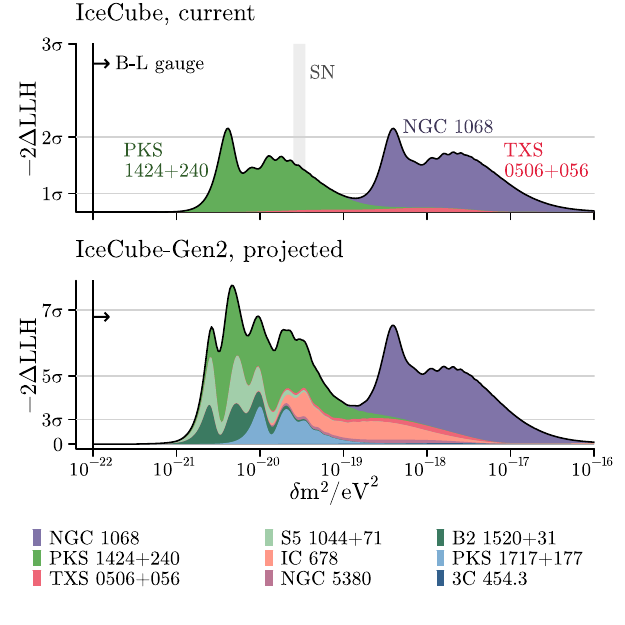}
  \caption{\textit{Top:} The log-likelihood ratio $\Delta$LLH of the combined analysis with the three currently significant sources (black) overlaid with that of each separately (color). The gray-shaded region indicates the $3\sigma$ region excluded by SN~1987~A.\textit{Bottom:} Projected sensitivity of IceCube Gen2, using nine astrophysical sources and assuming 8x statistics. Sensitivity from each source is stacked; the total is shown by the black curve.}\label{fig:delta_llh}
\end{figure}

The results of the sensitivity analysis are shown in~\Cref{fig:delta_llh}.
For TeV sources, the redshift of each source fixes the effective distance, $L_{\text{eff}}$, and thus the scale of the mass-splittings to which it is sensitive.
The spectral index of each source sets the distribution of its events over energy, which controls the width of the range of mass-splittings to which it is sensitive. 
Thus NGC~1068 and PKS~1424+240, which both have relatively soft spectra but have redshifts three orders apart, are sensitive to two very different, concentrated regions of parameter space.
Conversely, TXS~0506+056, which has a much softer spectra, is sensitive to a wide region, although its sensitivity is limited by its small best-fit event count (5 events).
% The peaks of the log-likelihood ratio correspond to the maximal predicted event disappearance in each energy bin. 
% The dips in the log-likelihood ratio are artifacts of the binning procedure.

At small values of the mass splitting ($\delta m^2\sim 10^{-21}\si\eV^2$), the sensitivity is dominated by PKS~1424+240, reaching $\sim 5\sigma$ for $\delta m^2\sim 10^{-20}\si\eV^2$.  
Interestingly, mass splittings on the same order have previously been explored using data from supernova SN1987A~\cite{Martinez-Soler:2021unz} (vertical grey-shaded region). 
For masses around $\delta m^2\sim 10^{-18}~\text{eV}^2$, the sensitivity is dominated by NGC~1068. 
The vertical line to the left of the plot indicates the left-edge of the region motivated by a $B-L$ gauge symmetry.

%At small values of the mass splitting, the pseudo-Dirac hypothesis, at these length-scales, reduces to the Standard Model.
%At large mass-splittings, the frequency of the pseudo-Dirac oscillations grows many times larger than the energy-resolution, and thus the oscillations average out to an unobservable normalization effect.

In supplemental analyses, we also consider whether a pseudo-Dirac signal would be recoverable. 
In~\Cref{sec:roundtrip} we calculate the likelihood of SM and pseudo-Dirac hypotheses, given data following a pseudo-Dirac prediction for two possible values of the equal mass-splitting.
We find that the maximum-likelihood point corresponds to the true value.
A pseudo-Dirac reality would also impact studies of source fluxes, by shifting the inferred spectral index with respect to the true value. 
We explore this possibility in in~\Cref{sec:spectral_index}, and find the effect could be as large as an 8\% shift.
This effect could also impact diffuse neutrino fluxes, since distortions to the source flux caused by pseudo-Dirac disappearances could accumulate.
Finally, in ~\Cref{sec:start_through} we explore the possibility of improving IceCube's current sensitivity by separating events into tracks which start within the detector volume, which have superior energy resolution, and those which traverse it. 
This possibility was not included in the main analysis because the starting event fraction is not publicly available.

In the lower part of~\Cref{fig:delta_llh}, we show the projected sensitivity of IceCube-Gen2.
For this analysis, we included all the sources which IceCube can currently identify with at least 1.1$\sigma$ local significance, and for which there exists a published redshift value; see~\Cref{table:sources}.
Additionally, we multiply all statistics by a factor of 8, as projected by Ref.~\cite{Song:2020nfh}.
We find that the combined sensitivity is well past 3$\sigma$ over a wide range of mass splittings.

\textbf{\emph{Conclusions.---}}

In this \textit{Letter}, we investigated IceCube's current and future sensitivity to the pseudo-Dirac neutrino mass scenario.
The combined analysis of the three most significant astrophysical sources observed by IceCube probes active-sterile mass-splittings in the range $10^{-20}\si\eV^2 < \delta m^2 < 10^{-16}\si\eV^2$, but its sensitivity is limited by statistics and poor energy resolution.
However, by including sources observed by IceCube with a significance larger than $1.1\sigma$, and assuming 8 times greater statistics, we found that IceCube-Gen2 will be able to explore a large range of masses with a significance over $5\sigma$. 

%In this \textit{Letter}, we showed that IceCube is sensitive to a pseudo-Dirac neutrino signal. 
%The combined analysis of the three most significant astrophysical sources observed by IceCube probes active-sterile mass-splittings in the range $10^{-20}\si\eV^2 < \delta m^2 < 10^{-16}\si\eV^2$ with a significance that can reach $5\sigma$.
%Including sources observed by IceCube with a significance larger than $1.1\sigma$, and assuming 8 times greater statistics, we found that IceCube-Gen2 will be able to explore a large range of masses with a significance over $9\sigma$.

Next-generation neutrino telescopes~\cite{IceCube-Gen2:2020qha,Adrian-Martinez:2016fdl,P-ONE:2020ljt,GRAND:2018iaj,Neronov:2016zou,Sasaki:2017zwd,Romero-Wolf:2020pzh,POEMMA:2020ykm} will unveil a constellation of neutrino sources, opening the possibility of exploring very long baseline neutrino physics. 
By combining many astrophysical sources, over a wide range of distances, we will gain access to a broad and hitherto unexplored range of active-sterile mass splittings. 
The expected signal, a dip in the neutrino spectra, due to the oscillation into the sterile state, must be observed in all the sources sharing a common $L_{\text{eff}}/E$. 
This signature is robust under uncertainties in astrophysical neutrino fluxes, which at this early point are significant.

The pseudo-Dirac hypothesis can also modify the flavor ratio of high-energy neutrinos at IceCube.
In presence of active-sterile oscillations, the flavor composition on Earth is expected to be different from the conventional (1:1:1) for a (1:2:0) source composition~\cite{Learned:1994wg}, depending on the flavor structure of the neutrino mass matrix~\eqref{eq:seesaw}.
Unfortunately, reconstructing the flavor triangle in this case will require both cascade and track events from the identified sources.
The poor angular resolution of the cascades ($\Delta\delta\sim 1^\circ$, as compared to $\Delta\delta\lesssim 0.1^\circ$ for tracks) makes this task rather difficult, but it may be possible with future neutrino telescopes.

This \textit{Letter} strongly motivates a full likelihood-based analysis by the IceCube collaboration. 
A more descriptive flux hypothesis would alter the significance of the source identification and the maximum likelihood parameters. This full likelihood analysis should be unbinned, consider background and signal simultaneously, and include a full treatment of detector systematics.
Only such a study would be able to unambiguously resolve the pseudo-Dirac nature of neutrinos.

%This \textit{Letter} strongly motivates a full likelihood-based analysis by the IceCube collaboration.
%A likelihood analysis is especially important because the significance of the identification of a source could depend on the flux hypothesis used.
%Thus the available public information given in Ref.~\cite{IceCube:2022der} is insufficient to derive constraints on this model.
%In fact, such a study will not only be able to resolve the pseudo-Dirac nature of neutrinos in an unambiguous way, but also potentially increase the significance of sub-threshold sources due to the improved spectral description.

\textbf{\textit{Acknowledgments ---}}
We thank Matthew Reece and William Thompson for useful comments. 
We also thank Jack Pairin for his artwork.
CAA and IMS are supported by the Faculty of Arts and Sciences of Harvard University and the Alfred P. Sloan Foundation.
KC is supported by the NSF Graduate Research Fellowship under Grant No.~2140743.
KSB is supported in part by the US Department of Energy grant No.~DE-SC~0016013.
BD is supported by the US Department of Energy grant No.~DE-SC~0017987 and by a URA VSP fellowship.
We thank the organizers of the NTN workshop at Fermilab in June 2022 for local hospitality where this work was initiated.

\bibliography{pseudo_dirac_astro.bib}

%%%%%%%%%%%%%%%%%%%%%%%%%%%%%%%%%%%%%%%%
% supplemental materials
\clearpage
\pagebreak
\appendix
\onecolumngrid
\ifx \standalonesupplemental\undefined
\setcounter{page}{1}
\counterwithin{figure}{section}
\numberwithin{equation}{section}
\fi
\renewcommand{\thepage}{Supplemental Methods and Tables --- S\arabic{page}}
\renewcommand{\figurename}{SUPPL. FIG.}
\renewcommand{\tablename}{SUPPL. TABLE}
\renewcommand{\theequation}{A\arabic{equation}}
\clearpage
\begin{center}
\textbf{\large Supplemental Material}
\end{center}
%%%%%%%%%%%%%%%%%%%%%%%%%%%%%%%%%%%%%%%%

\section{Candidate astrophysical sources}

\begin{table}[h]
\setlength{\tabcolsep}{1em}
\centering
\begin{tabular}{l l l l l l l}
\toprule
Source & Source Type & $-\log_{10}p_{\rm local}$ & $\hat{n}_s$ & $\hat{\gamma}$ & $z$ & $z$ Ref. \\
\hline
NGC~1068 & SBG/AGN & 7.0 & 79 & 3.2 & $0.0038 \pm 0.00001$ & \cite{Meyer:2004hr} \\
PKS~1424+240 & BLL & 4.0 & 77 & 3.5 & $0.6047 \pm 0.1$ & \cite{Paiano:2017pol} \\
TXS~0506+056 & BLL/FSRQ & 3.6 & 5 & 2.0 & $0.3365 \pm 0.001$ & \cite{Paiano:2018qeq} \\ \hline
S5~1044+71 & FSRQ & 1.3 & 45 & 4.3 & 1.1500 & \cite{1995ApJS} \\ %refId0} \\
IC 678 & GAL & 0.9 & 22 & 3.1 & $0.04799 \pm 0.00002$ & \cite{SDSS:2016ear}\\
NGC 5380 & GAL & 0.9 & 4 & 2.4 & $0.010584 \pm 0.000077$ & \cite{Huchra_Geller_Corwin_1995}\\
B2 1520+31 & FSRQ & 1.0 & 35 & 4.3 & $1.48875 \pm 0.00025$ & \cite{SDSS:2016ear} \\
PKS 1717+177 & BLL & 1.0 & 34 & 4.3 & 0.137 & \cite{Sowards_Emmerd_2005} \\ %refId0}\\
3C 454.3 & FSRQ & 1.2 & 1 & 1.5 & 0.859 & \cite{10.1093/mnras/sty2834} \\
\end{tabular}
\caption{The nine astrophysical sources considered in this work. The first three are the most significant. The maximum likelihood values $-\log_{10} p_{\rm local}, \hat{n}_s, \hat{\gamma}$ are copied from Ref.~\cite{IceCube:2022der}. The redshifts are obtained from different sources, as listed in the last column. The source GB6~J1542+6129, which was identified by Ref.~\cite{IceCube:2022der} with local significance of 2.2$\sigma$, was not included in this analysis, as its redshift is poorly constrained.}
\label{table:sources}
\end{table}

\section{Energy resolution and sensitivity
}
\label{sec:energyres}

\begin{figure}[htb]
  \centering
  \includegraphics[width=\textwidth]{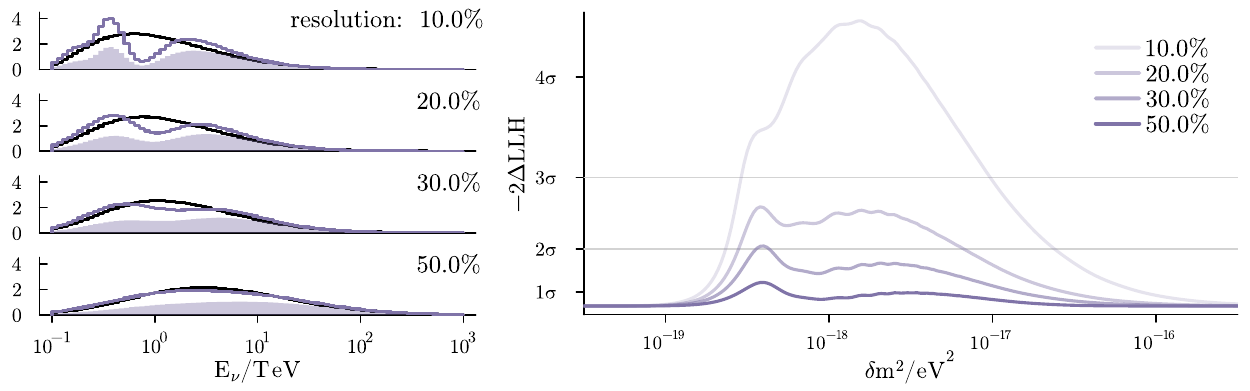}
  \caption{Event distributions (\emph{left}) and sensitivities (\emph{right}) for different values of the log-energy resolution, using NGC~1068 as the source.}\label{fig:res_scan}
\end{figure}

The energy resolution of the IceCube detector significantly affects the ability to resolve a dip in the event distribution. 
In this section we demonstrate this effect.
We model the probability of an event with given reconstructed log-energy having a particular true value as a Gaussian with variable width. 
On the left side of  ~\Cref{fig:res_scan}, we plot the event distributions at $\delta m^2 = 10^{-17.8} \text{eV}^2$ for different values of this energy resolution, while on the right side we plot the corresponding sensitivity to the pseudo-Dirac parameter space.
As the resolution improves, the pseudo-Dirac disappearance effect increases in clarity, and the sensitivity leaps upward.

The Gaussian model used here is only an approximation to the reconstructed energy distribution. Generally, the true energy of an event can be significantly underestimated by reconstruction algorithms when the track length extends far beyond the detector confines.
This underestimation effect results in a true energy probability distribution with a long asymmetric tail.
Additionally, this effect is enhanced for higher-energy events, which travel much further in the ice.

\section{Sensitivity to distinct mass splittings}\label{sec:severalmass}

Although most of this analysis focused on the subset of parameter space in which all three pseudo-Dirac mass splittings are equal, we have also performed a scan for the sensitivity to distinct mass splittings. 
Because this work focuses on track-like events in IceCube, whose direction can be reconstructed to sub-degree precision, the mass splittings $\delta m^2_2, \delta m^2_3$ are much more significant than $\delta m^2_1$. This is due to the fact that track-like events are predominantly produced by muon neutrinos, and the mass states $\nu_2, \nu_3$ are more likely to be measured in the flavor basis as $\nu_\mu$ than $\nu_1$.

\begin{figure}[htb]
  \centering
  \includegraphics[width=0.6\textwidth]{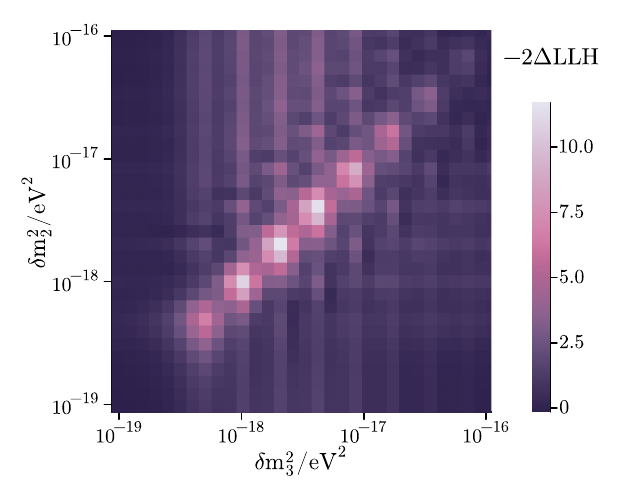}
  \caption{Sensitivity to the pseudo-Dirac parameter space with distinct mass splittings. Here we have varied $\delta m^2_{3}$ and $\delta m^2_{2}$, with fixed $\delta m^2_1 = 0.0$.}\label{fig:gridscan}
\end{figure}

In ~\Cref{fig:gridscan} we plot the test statistic $-2 \Delta \text{LLH}$, corresponding to the likelihood ratio of the pseudo-Dirac hypothesis to the SM, calculated for data simulated according to the SM prediction for NGC~1068. In this section we also assume near-perfect energy reconstruction. The sensitivity is maximized at the center of the diagonal, where a maximum number of neutrinos in both mass states $\nu_2, \nu_3$ are oscillating into their sterile counterparts and disappearing.

\section{Injection and recovery of true pseudo-Dirac parameters}
\label{sec:roundtrip}
\begin{figure}[htb]
  \centering
  \includegraphics[width=0.6\textwidth]{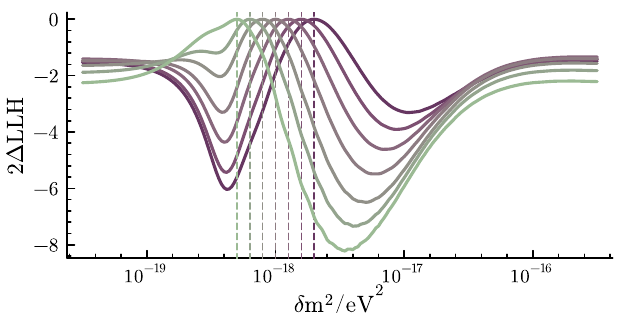}
  \caption{$\text{LLH}(\text{data}|\delta m^2) - \text{LLH}(\text{data}|\delta m^2_{\text{true}})$, where the data was simulated as pseudo-Dirac neutrinos from NGC~1068, and 30\% log-energy resolution. The injected values of the mass splittings, $\delta m^2_{\text{true}}$ are plotted as dashed vertical lines.}\label{fig:roundtrip}
\end{figure}

In this section, we consider whether a maximum likelihood fit would be able to recover the true pseudo-Dirac parameter values. 
Using the event distribution of the source NGC 1068, we generate an event distribution according to an injected uniform pseudo-Dirac mass splitting, $\delta m^2_{\text{true}}$ and then calculate the likelihood of this distribution under a different pseudo-Dirac hypothesis, $\delta m^2$. The test statistic 2$\Delta\text{LLH}$ is plotted for a spectrum of injected $\delta m^2_{\text{true}}$ in~\Cref{fig:roundtrip}.
%In ~\Cref{fig:roundtrip} we show the log-likelihood as a function of different values of a uniform pseudo-Dirac mass splitting. %The log-likelihood is normalized by subtracting out the log-likelihood of the best-fit value.
% Using the event distribution of the source NGC 1068, we consider two different injected values for the uniform pseudo-Dirac mass splitting. : $\delta m^2 = 10^{-17.72} ~\text{eV}^2$, which is the point with maximum sensitivity in ~\Cref{fig:delta_llh} (purple), and a second point with lower sensitivity (pink).
%$\delta m^2 = 10^{-16.83} ~\text{eV}^2$, 

In all cases the likelihood difference is maximized at the injected value.
The likelihood of the SM hypothesis can be equated with the limit $\delta m^2 = 0$ in this plot, i.e. the limiting value of the left-hand side. 
% As expected, we are able to reject alternate hypotheses with greater significance when we inject the value to which we had greater sensitivity.

\section{Standard Model fits to pseudo-Dirac reality
}
\label{sec:spectral_index}
\begin{figure}[htb]
  \centering
  \includegraphics[width=\textwidth]{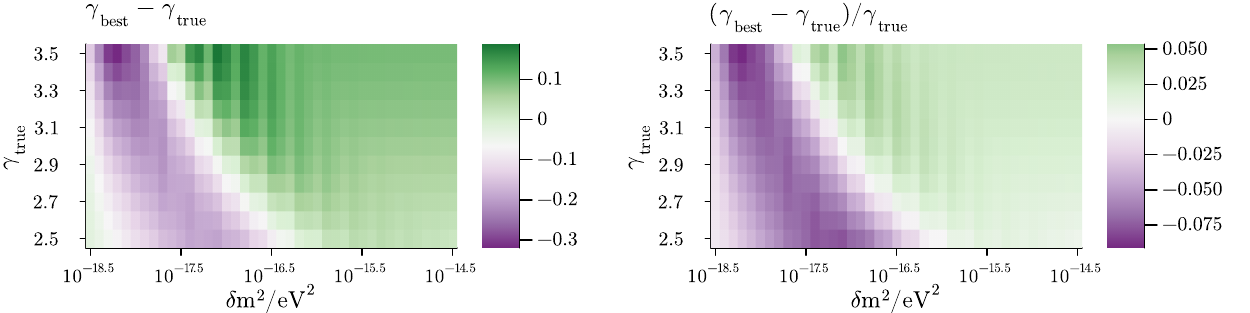}
  \caption{Absolute (\emph{left}) and relative (\emph{right}) error of a fit of NGC 1068's spectral index, where the data has been simulated according to a pseudo-Dirac prediction, but the fit is performed assuming a SM hypothesis.}
  \label{fig:fit_spectral_index}
\end{figure}

We also consider how a maximum likelihood fit for source flux parameters (normalization and spectral index) which assumed a Standard Model hypothesis would perform if the data was truly pseudo-Dirac. 
In ~\Cref{fig:fit_spectral_index} we plot the absolute and relative errors on a SM fit of NGC~1068's spectral index, as a function of the true spectral index $\gamma_{\text{true}}$ and the value of the uniform pseudo-Dirac mass splitting $\delta m^2$. 
In this study we again assume near-perfect energy reconstruction.

In our binned likelihood analysis, each value of $\delta m^2$ results in a deficit of events in some combination of energy bins. Depending on the value of the spectral index, the SM hypothesis may or may not predict many events in those bins, which means the deficit may or may not be significant. If a significant number of events go missing in low energy bins, the SM hypothesis will overfit the spectral index, preferring a flatter distribution. Conversely, if a significant number of events go missing in high energy bins, the spectral index best fit value will underestimate the true value. 

\section{Sensitivity for starting and through-going tracks
}
\label{sec:start_through}

Reconstruction algorithms struggle to estimate how much energy a muon track deposited outside of the detector volume, resulting in systematic underestimation of the true event energies \cite{Palladino_2018, IceCube:2013dkx}. 
This problem is less significant when considering only the sub-sample of starting tracks, whose interaction vertex is contained within the detector.
Reconstruction algorithms designed specifically for starting tracks can therefore achieve significantly better performance, especially at high energies.

\begin{figure}[htb]
  \centering
  \includegraphics[width=\textwidth]{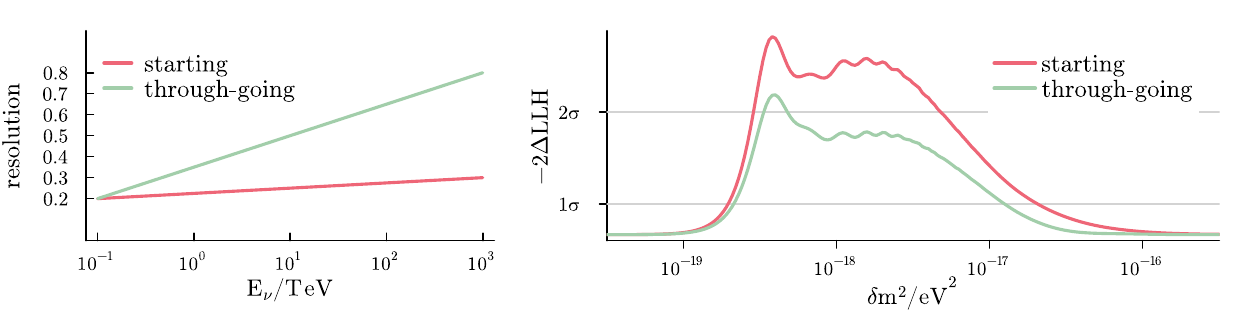}
  \caption{\emph{Left:} The linear models for the energy resolution of starting and through-going events as a function of energy. \emph{Right:} Sensitivity to the pseudo-Dirac parameter space, assuming all events have starting or through-going resolution.} 
  \label{fig:sens_start_through}
\end{figure}

In this subsection we consider the sensitivity that can be achieved if all the event have through-going or starting resolution. 
We model each resolution as a simple linear function~\cite{Alfonso2022}, and calculate the sensitivity using NGC~1068 as the source. 
The models and the resulting sensitivity curves are shown in~\Cref{fig:sens_start_through}.
In the case in which we assume all events have starting-quality resolution, the sensitivity performance is similar to the 20\% curve in~\Cref{fig:res_scan}, while the through-going case is more similar to ~30\%. 

\end{document}